\pdfoutput=1
\documentclass[twocolumn]{article}
\usepackage{graphicx}
\graphicspath{{../figures/}{figures/}}
\usepackage{authblk}
\usepackage{caption}
\usepackage{color}
\usepackage{hyperref}
\usepackage{listings}
\usepackage{pgfplots}
\usepackage{tikz}
\usepackage{seqsplit}
\usepackage{longtable}
\usepgfplotslibrary{statistics}
\pgfplotsset{compat=1.5}

\definecolor{dkgreen}{rgb}{0,0.6,0}
\definecolor{gray}{rgb}{0.5,0.5,0.5}
\definecolor{mauve}{rgb}{0.58,0,0.82}

\lstdefinelanguage{Solidity}{
  keywords={function, sstore, ":=", "->", byte, assembly, external, returns, uint, return, sload},
  keywordstyle=\color{gray}\bfseries,
  identifierstyle=\color{black},
  sensitive=false,
  comment=[l]{//},
  morecomment=[s]{/*}{*/},
  commentstyle=\color{purple}\ttfamily,
  stringstyle=\color{red}\ttfamily,
  morestring=[b]',
  morestring=[b]"
}

\newcommand{\fuzzer}{SolSmith}
\begin{document}
\title{Finding and Understanding Miscompilation Bugs in the Solidity Compiler}
\author[]{Bhargava Shastry}
\affil[]{Ethereum Foundation\\bhargava.shastry@ethereum.org}
\date{}
\setcounter{Maxaffil}{0}
\renewcommand\Affilfont{\itshape\small}
\date{}
\maketitle
\begin{abstract}
Smart contract compilers are critical to ensuring the correctness of public blockchains whose defining characteristics are open-source and immutable code.
We created \fuzzer{}, a semantics-aware differential fuzz testing tool, to improve the quality of the Solidity compiler---the most popular compiler for the Ethereum blockchain---and spent over three years finding compiler defects that produce incorrect code.
We call these defects miscompilation bugs.
During this time period, we have discovered 25 miscompilation bugs that went unnoticed, some for multiple years.

Our first contribution is to make compiler testing more rigorous.
\fuzzer{} achieves this goal by generating valid test programs that are likely to stress test code generation and optimization components.
This helps \fuzzer{} find bugs missed during routine testing that could potentially have serious implications for smart contracts and their users.
Our second contribution is a qualitative and quantitative analysis of miscompilation bugs that we found in the Solidity compiler.
We classify miscompilation bugs found by \fuzzer{} based on their nature, root-causes, and impact on end-users.
This sheds light on some pitfalls of optimizing compilers.

\end{abstract}

\section{Introduction}
Smart contracts are programs that codify the terms of an agreement between contracting parties.
They are written in domain-specific programming languages like Solidity and are executed on the Ethereum blockchain.
The compiler is responsible for translating the high-level smart contract code written in Solidity into bytecode that can be executed on the Ethereum Virtual Machine (EVM).
It is no surprise that compilers for smart contracts, like any complex software system, contain bugs.
In this paper, we look specifically at miscompilation bugs: compiler defects that result in incorrect code generation.

Miscompilations are dangerous in general because they may go unnoticed unless the compiled code has been thoroughly tested.
Miscompilations of smart contracts deployed on the Ethereum blockchain are particularly problematic because (1) deployed code is immutable (bugs persist); (2) they may alter the blockchain's persistent state (side effects of bugs persist); and (3) side-effects routinely include financial transactions (sending/receiving Ether or other ERC20 tokens).

In our experience, miscompilations occur due to two main reasons: incorrect optimization and incorrect translation.
Miscompilations due to incorrect optimizations occur for the following reasons.
First, if optimization correctness checks are either inadequate or not thoroughly tested such that in most cases the optimization produces the desired output with the exception of corner cases.
Second, if a formally verified optimization rule is correct in theory but incorrect in practice (e.g., function side-effects are not taken into account leading to more aggressive optimization than permissible).
Third, if wrong assumptions are made on the code to be optimized (e.g., may conditions are assumed to be must).
Miscompilations due to incorrect translation occur either due to implementation flaws or inconsistent cleanup of typed data.

Verifying the correctness of the compiler is challenging, and traditional testing methods, such as writing unit and end-to-end tests, have limited effectiveness because they do not fully tease out compiler output for diverse programs.
Blackbox fuzz testing---simply generating random input---is of little value when it comes to compiler testing if the goal is to find miscompilation bugs.
This is because the test input needs to trigger code generation to be able to test for correct compilation and this is highly unlikely to occur when the input to the compiler is a random sequence of bytes.
In order for a test input to cause the compiler to emit code, it must do something within the semantic rules of the programming language.
Otherwise, the compiler errors.
To trigger code generation, some form of semantic awareness is required.
One of the forms of whitebox fuzz testing that have been proposed to test compilers involve generating programs based on the language's context-free grammar~\cite{godefroid}.
Although whitebox fuzzing techniques perform slightly better than blackbox approaches, they rarely trigger code generation because of semantic errors such as mismatched types, incorrect statement placement etc.
Our hypothesis is that semantic expressiveness of test input is correlated with its bug finding potential.

We create~\fuzzer{} for finding miscompilation bugs in the Solidity compiler, the most popular smart contract compiler for the Ethereum blockchain.
We discover two classes of miscompilation bugs: (1) Incorrect code due to faulty program translation; (2) Incorrect code due to faulty program optimization.
\fuzzer{} pseudo-randomly generates semantically correct Solidity programs and invokes the compiler against them.
The testing is differential in nature.
The compiler is invoked twice for the same input program but in non-identical compiler configurations, namely the control configuration (C) and the experiment configuration (E).
The configurations are chosen such that they differ in exactly one setting that controls a specific function of the compiler that is to be tested.
Each invocation produces bytecode that is run on a production EVM in an identical state.
The two runs are traced at runtime to obtain execution traces corresponding to configurations~\textbf{C} and~\textbf{E}.
Normally, both runs should result in an identical execution trace because the compiler is expected to preserve program semantics during compilation.
If traces are non-identical, the input program and the compiler configurations are saved to validate if the divergence is due to a miscompilation bug or not.

We find miscompilation bugs of class 2 (incorrect optimization) by comparing the runtime of unoptimized and optimized invocations of the compiler.
Finding miscompilation bugs of class 1 (incorrect translation) is tricky because we need a suitable baseline implementation.
To do this, we leverage existing alternative (semantics preserving) implementations of the code generator (legacy and IR-based); one is chosen as the baseline while the other as experiment configuration.
Using the proposed method, we have identified 25 miscompilation bugs in the compiler that have been patched.
Because random testing lends itself well to automation, we have been using it as part of the development life-cycle of the compiler to identify bugs as early as possible.

Listing~\ref{lstbug} illustrates a miscompilation bug for a semantically valid program written in the Yul IR language.
The Yul program contains a function~\texttt{byteIndexedBy(n)} that returns the~\texttt{nth} byte of the constant 256-bit (32-byte) value~\texttt{31} indexed by its input argument~\texttt{n}.
The output of this function invoked with~\texttt{1} as the input parameter is written to the EVM blockchain's persistent storage slot zero in the subsequent line.
Since the 256-bit value~\texttt{31} may be represented by~\texttt{0000...1f} (first 31 most significant bytes zeroes, the least MSB non-zero), and indexing starts from the most significant byte (index zero), the function call must return the value zero.
However, this program compiled using the Solidity compiler version~\textbf{0.5.5} would return one.
The miscompilation occurred due to an optimization rule that transforms byte accesses of constant values into a mask (\textbf{and(value, 0xff)}).
The rule targeted the wrong input argument for optimization: instead of targeting~\textbf{byte(31, X)} (\textbf{X} being the constant expression) for optimization, the rule targeted~\textbf{byte(X, 31)} and transformed it into~\textbf{and(X, 0xff)}
Therefore,~\textbf{byte(n, 31)} is evaluated as~\textbf{byte(31, n)} yielding the value one instead of the correct value of zero.

\begin{table*}[t]
\begin{tabular}{cc}
\begin{minipage}{.4\textwidth}
\begin{lstlisting}[caption=Yul code snippet before optimization., label=lstbug]
{
  // Function that outputs the byte indexed by
  // input n of the constant 256-bit value 31.
  // 0 indexes the most significant byte of 31 (= 0).
  // 31 indexes the least significant byte of 31 (= 31).
  function byteIndexedBy(n) -> o
  {
    // Optimization rule bug: Swaps order of arguments
    // Assumes 32'nd most significant byte of n.
    o := byte(n, 31)
  }
  // Store function output to slot zero
  sstore(0, byteIndexedBy(1))
}
\end{lstlisting}
\end{minipage}
\begin{minipage}{.45\textwidth}
\begin{lstlisting}[caption=Yul code snippet after optimization., label=lstbugoptimized]
{
  // Expected: sstore(0, 0) because the second MSB of
  // 31 is zero.
  // Actual result: sstore(0, 1) because the optimizer
  // incorrectly constant folds and optimizes for the
  // least significant byte of the constant 1.
  sstore(0, 1)
}
\end{lstlisting}
\end{minipage}
\end{tabular}
\end{table*}

This paper makes two contributions.
Our first contribution is to make compiler testing more rigorous by generating expressive source programs and executing their compilation output differentially.
~\fuzzer{} has found 25 miscompilation bugs in over three years of testing, some of which were discovered during continuous integration i.e., while a pull request was being reviewed.
Our second contribution is a qualitative analysis of miscompilation bugs found by~\fuzzer{}: Where are they present? What is their root-cause? How do they affect end-users?
We find that (1) close to two thirds of miscompilation bugs occur due to incorrect optimization, and the rest due to incorrect code translation; (2) miscompilation bugs due to incorrect optimization mostly occur because of underconstrained application of an optimization rule; (3) miscompilation bugs found by~\fuzzer{} have not been externally reported to us leading us to believe our effort is proactive.

\section{Related Work}
Broadly speaking, there are two directions that have been taken to find miscompilation bugs: (1) Random generational testing; (2) Equivalence modulo inputs.

\subsection{Random generational testing}
jsfunfuzz~\cite{jsfunfuzz} is a JavaScript fuzzer that finds miscompilation bugs by targeting Mozilla's non-standard JavaScript decompiler interface.
A randomly generated valid JavaScript program is first compiled, then decompiled to obtain the original source, and finally compiled again to check for syntax errors.
This helps discover bugs in the decompiler that would alter the syntax of the original code.
Similarly, consistency bugs are reported by the author by checking if two ``round-trips'' (compilation plus decompilation) produce an identical result.
Our work uses a different---arguably more important---notion of a miscompilation bug to mean compiler bugs that affect compilation rather than decompilation.

Grammar-based whitebox fuzzing~\cite{godefroid} uses a constraint-solver to generate JavaScript programs to increase the test coverage of the JS interpreter; it is not evaluated whether the proposed approach affects bug-finding ability or if it actually discovers miscompilation bugs as defined by our work.
JS programs are generated using the context-free grammar and the constraint-solver is used to prune generations such that they stay syntactically valid.
In contrast, \fuzzer{} generates context-sensitive programs that respect language semantics (e.g., a \textit{break} statement may only be inserted inside a loop statement, type checking on variables must pass etc.).
More importantly, \fuzzer{} is tailored to discover miscompilation bugs.

CSmith~\cite{csmith} is a differential testing tool targeted at C compilers, and closest to our work.
It generates test programs that respect the C language semantics (e.g., no undefined behavior, type safety respected etc.) and finds bugs by comparing the runtimes of binaries generated by multiple compiler implementations.
The scope of that work is to find bugs in \textit{any} C compiler unlike our aim of finding bugs in the Solidity compiler.
The biggest difference between CSmith and \fuzzer{} is that the latter benefits from domain-knowledge: generations and mutations that actually test specific optimization steps are implemented.
For example, to test the redundant write-to-memory eliminator (that removes redundant writes to memory as the name suggests), writes to memory are generated that are later either read-from or not-read-from testing whether the test cases containing non-redundant write and redundant writes respectively do not lead to incorrect code generation i.e., a non-redundant write being wrongly removed.

Equivalent Modulo Inputs (EMI)~\cite{emi} is a compiler testing method that involves feeding multiple semantics-preserving programs to the compiler and checking if they produce the same output to a given set of inputs.
This is complementary to random differential testing and may be applied to the inputs generated by \fuzzer{}.
Since EMI does not perform differential testing of multiple implementations, it relies quite heavily on the correctness of semantics-preserving mutations.
Although certain mutations are easy to implement (e.g., adding/removing dead code), the mutations that would be useful to find bugs in specific optimization routines (e.g., redundancy elimination) require precise program analysis; lack of precision would lead to false positives.

\section{Background}
In this section, we provide a brief overview of the Solidity programming language and the compiler implementation.

\subsection{Solidity Programming Language}
\label{para:sol}

Solidity is a high-level programming language~\cite{soliditydocs} that is used to develop programs---so-called smart contracts---to be deployed on the Ethereum Virtual Machine (EVM).
The compilation entry point is a contract or a library.
A contract may contain one of more functions that may accept Ether, EVM's native currency.
There are both statically and dynamically typed variables in Solidity.
Static types include integers (both signed and unsigned), fixed width bytes (width of 1 upto 32 bytes), boolean, address, and statically sized arrays of static types.
An address holds a 20-byte key that could be used to index either a deployed contract or a wallet. 
Dynamic types include variable width bytes or strings, structs, mappings and dynamically sized arrays.
A mapping is a key-value type.

\subsection{Solidity Compiler}
\label{sec:solcom}

The Solidity compiler may be viewed as a conjunction of a front-end (FE), middle-end (ME), and a back-end (BE).
Until the development of the Yul intermediate representation, the Solidity compiler only had a front-end and a back-end.
With the production release of the Yul based code generation engine~\cite{soliditydocs}, the Solidity compiler supports two compilation pipelines: legacy and IR-based.
The two pipelines share the FE and BE.
The legacy pipeline comprises the FE and an EVM bytecode based BE.
The IR-based pipeline comprises the FE, the IR-based ME, and the BE.

The legacy code generator translates Solidity programs directly to EVM bytecode.
The Intermediate Representation (IR) based code generator is based on an IR called Yul.
The IR-based code generator translates Solidity programs to Yul, then optimizes it retaining the optimized output in Yul form, and finally assembles optimized Yul to EVM bytecode.

\paragraph{Front End}

The front end parses Solidity source code into an abstract syntax tree (AST) and performs semantic analysis on it.
Semantic analysis includes name resolution, type checking, and control-flow checks; it annotates the AST with the information that code generation depends on.
Both compilation pipelines consume the same annotated AST, which makes the front end a shared component of the differential test setup.

\paragraph{Middle End}

Yul is an intermediate representation (IR) used for ease of optimizing Solidity programs.
It is untyped, data being 256-bit in size.
Because Yul contains control-flow constructs like loops, switch statements it permits a higher-level view than EVM bytecode can afford yet is close enough to EVM to ease assembly. 

The Yul optimizer can perform optimizations that are not possible by the byte code optimizer.
Since the Yul IR does not permit arbitrary jumps, function side-effects may be computed.
This aids in, for example, re-ordering function calls or removing function calls entirely (e.g., if the result of a single output function call without side-effects is multiplied by zero).

\paragraph{Back End}

The bytecode optimizer optimizes EVM bytecode that is output by either of the two code generation engines.

\subsection{Ethereum Virtual Machine}
\label{para:evm}

The Ethereum Virtual Machine (EVM) is a distributed computing system that maintains a shared state and works on the principle of consensus.
The shared state comprises a mapping of accounts that are referenced via 160-bit \textit{addresses} to their account state.
An account state comprises code, storage, nonce, and a balance.
Nonce and balance are machine words~\footnote{An EVM machine word is 256-bits in size} used to prevent replay attacks and maintain the Ether held by the account.
An account may be either be externally owned (EOA) or a smart contract.
An EOA is managed by a public-private key pair but contains no code as such.
A smart contract contains code (sequence of bytes of EVM bytecode) and can access non-volatile \textit{storage}.

At any given point, the EVM has a singular state.
State transitions are governed by \textit{transactions}.
Transactions are of two types: \textit{create}, and \textit{call}.
The former deploys a smart contract (bytecode) to a new address on the Ethereum blockchain; the latter calls another account.
A call is accompanied by transferred balance, \textit{gas} that limits the call's access to computational resources, and data (sequence of bytes).

\section{\fuzzer{}}
\begin{figure}
\centering
\includegraphics[width=0.3\textwidth, angle=0]{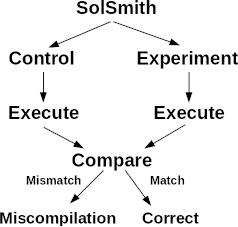}
\caption{Finding miscompilation bugs using differential fuzz testing}
\label{fig:overview}
\end{figure}

One of the key requirements to find miscompilation bugs is compilability: test programs must compile without errors.
This rules out black-box fuzzers that do not respect language semantics because although we may supply these fuzzers with a sampling of valid input (existing test cases), it is highly unlikely that they mutate them in a semantics preserving manner.
Thus, we set out to design a white-box fuzzer called \fuzzer{} that understands Solidity syntax as well as semantics.

Successful compilation is not sufficient to find miscompilation bugs because we need a reference for what may be deemed correct compilation.
Since there is little control over the nature of programs generated by \fuzzer{}, it is impossible to find miscompilations simply by comparing the output of the compiler for a given program with the byte code that the input program is expected to produce.
Therefore, we adopt a runtime testing approach.

\subsection{Differential Fuzz Testing}
A miscompilation bug is flagged if two distinct semantics preserving code generation pipelines produce an executable whose runtime semantics differ.
We call the baseline the control group, and the test subject the experiment group.
Runtime semantics of a compiled smart contract include stateful changes it makes, the returned result of a call initiated against it, and the status of the initiated call.

To find miscompilation bugs in the end-to-end code generation, we use Solidity compiler's legacy code generator implementation as the control group and its new Yul IR-based code generator implementation as the experiment group.
To find miscompilation bugs in the Yul optimizer, we use unoptimized code as the control group and the optimized code as the experiment group.

\subsection{Design Goals}

\fuzzer{} has three main design goals.
First, every test program must be semantically valid that has a single interpretation.
This imposes two requirements: (1) the test program contains no syntax or typing errors; (2) documented divergences in control and experiment groups that preclude unique interpretation must be avoided.
A unique interpretation means that the generated byte code has well-defined side effects and performs the same computation.

Second, generated test programs capture the expressiveness of the Solidity programming language.
This means that test programs make use of as many language constructs as possible.
\fuzzer{} creates programs with the following features:
\begin{itemize}
\item contracts and library definitions
\item function definitions
\item state variables, immutables, and constants
\item most kinds of Solidity expressions and statements
\item most kinds of Yul (inline assembly) expressions and statements
\item control flow statements such as \texttt{if/else if/else}, loops, \texttt{break}, \texttt{continue}, \texttt{return}
\item typed variables covering most kinds of Solidity types
\item user-defined types such as \texttt{struct}, \texttt{enum}
\end{itemize}

Third, test programs should be constructed incrementally from other test programs.
This means that it should be possible to slightly alter an existing (semantically valid) test program to create another (semantically valid) test program.
Our hypothesis is that incremental test generation (via mutations) is correlated with bug-finding ability.

\subsection{Program Generation and Mutation}

\fuzzer{} generates test programs on the basis on language grammar.
It synthesizes test cases in a top-down manner.
Each test case to the compiler is a sequence of Solidity files, each Solidity file may contain zero or more functions, contracts, libraries, and so on and so forth.
A function body is a statement block i.e., a sequence of statements.
\fuzzer{} supports most statement types defined by the Solidity programming language.
A contract body may contain state variables, constants, immutable variables, and zero or more functions that may accept Ether (payable functions)~\footnote{A full grammar specification supported by \fuzzer{} may be found at \url{https://github.com/ethereum/solidity/blob/develop/docs/grammar/SolidityParser.g4}}.
In addition, a contract may define a~\textit{constructor} function that constructs the contract for later use.
We generate simple forms of inheritance: a contract may derive from another~\textit{base} contract, overriding one or more functions.
We also synthesize inline assembly statements based on an independent specification of the Yul syntax.

In order to generate error-free test programs, \fuzzer{} defines a type system for the Solidity programming language.
The type system is capable of supporting aspects of an object-oriented language (objects, methods, inheritance) by maintaining a record of objects and their methods and typed instance variables.
\fuzzer{} maintains a~\textit{global state} in source file scope and multiple local states that refer to records in one of the following scopes: contract, function, and inline assembly block.
The global and local states are updated as we generate a new entity.
For example, the generation of a contract creates a contract state that will hold entities in its namespace (state variables, user-defined types, constants) that will be generated in the scope of that contract.

\fuzzer{} commences test program generation by creating a set of Solidity test files.
A test file is created as follows.
\begin{enumerate}
\item \fuzzer{} randomly selects a term from the top-level language grammar and invokes its production.
It consults a probability table in order to select a term from the set of all top-level terms.
Each term in the language grammar has an associated (user-controlled) probability of being generated.
The production is discarded if certain criteria are not met.
For example, production of the \texttt{using...for} statement---that declares functional operators on types---is discarded if no function is available for use.
If a production is discarded, \fuzzer{} selects another top-level term uniformly at random.
\fuzzer{} recurses should the selected term contain itself e.g., \texttt{block} statement within an outer block.
Otherwise, it visits the production rule's right-hand-side term.
\item If the selected production has dependencies e.g., function call that requires input arguments, the dependent term is either looked up from the scoped record (e.g., a variable reference in scope) or produced on-the-fly (e.g., literal expression).
\item Since Solidity enforces types rather strictly, a typed expression may only be formed from sub-expressions of the same type or another type that is acceptable by the Solidity type checker.
To produce type conformant expressions, \fuzzer{} annotates expressions with types (see Section~\ref{sec:filters}).
\item Should the chosen term be an inline assembly block, \fuzzer{} starts iterating on a top-level inline assembly terms, the grammar of which is independently specified.
As with Solidity statements, it recurses into or discards productions as necessary.
Compared to Solidity, inline assembly is untyped and not burdened with as much contextual annotation (e.g., function visibility, mutability) which makes its production rather straightforward.
\item The maximum number of test files per test case is user-controlled.
\fuzzer{} nears completion of a test case production once a randomly chosen number of test files have been generated.
It writes the test configuration before finalizing a test case.
The configuration includes parameters that define the control and experiment groups that are to be differentially tested.
This configuration is parsed by the fuzzer test harness which then executes the test accordingly.
\end{enumerate}

Once a test case is generated by \fuzzer{}, it is parsed by the fuzzer harness.
The fuzzer harness parses the contracts, libraries, and their functions from the test case.
The test configuration defines the control and experiment groups of the differential test.
The test harness executes the following steps for each of the two groups.

\begin{enumerate}
\item It chooses a contract uniquely at random, and deploys it to the blockchain.
\item It chooses one of the contract's functions uniquely at random and invokes it with typed arguements to that function which have also been generated randomly.
\item The output and status code of the function invocation, and the state of the blockchain are preserved for comparison later on.
The output is a sequence of bytes whose length is specified by the function output argument list.
The status code is an enum that indicate whether the call was successful or not, and if not the reason for failure.
The state of the blockchain comprises a string representation of the contract storage, the sequence of calls initiated by the called function, and log data.
\end{enumerate}

A miscompilation is flagged if the control and experiment groups diverge in at least one of the compared indicators: output, status code, and blockchain state.
Flagged miscompilations are inspected manually to sort out false positives.

\subsection{Filters to Reduce False Positives}
\label{sec:filters}


\begin{table}
\begin{tabular}{p{2cm} | p{2.5cm} | l}
Problem & Solution & Enforcement \\
\hline
typing errors & type system; lookup table of permissible types & static \\
scoping errors & type system & static \\
unspecified eval. order & disallow expressions with unspecified eval. order & static \\
large memory expansion & user-defined upper-bound & dynamic \\
code comparison & disallow & static \\
\end{tabular}
\caption{Summary of \fuzzer{}'s strategies for avoiding false positives and code generation errors.}
\label{tbl:fpengineering}
\end{table}

Table~\ref{tbl:fpengineering} summarizes the strategies employed by \fuzzer{} to avoid false positives and code generation errors.
Most strategies are enforced statically, i.e., during test program generation; the bound on memory expansion is enforced dynamically, i.e., while the compiled test program is executed.

\paragraph{Type and Memory Safety}

To rule out typing errors, \fuzzer{}'s type system tracks the type of every generated expression.
In addition, it consults a lookup table to choose which expression kinds are suitable for a given type.
The lookup table maps a given type to a list of expression lookup tables for that type and their probability of selection.
An expression lookup table maps a given expression kind to a set of probabilities for its operand types.
For example, Solidity's 256-bit unsigned integer type (\texttt{uint256}) can hold arithmetic expressions composed from other unsigned integer types of various widths; the lookup table for \texttt{uint256} therefore admits, among others, arithmetic expressions whose operands are unsigned integers of at most 256 bits, and encodes the probability with which each operand width is selected.
The same mechanism rules out scoping errors: only identifiers recorded in one of the currently visible scopes may be referenced by a generated expression.
To keep test execution within practical resource limits, memory accesses in generated inline assembly are bounded by a user-defined upper limit on memory expansion that is enforced at runtime.

\paragraph{Unspecified evaluation order}

The Solidity language does not specify the order in which sub-expressions of an expression are evaluated.
The two code generation pipelines are therefore free to evaluate sub-expressions in different orders, and they do so in practice.
A program whose observable behavior depends on the evaluation order would cause the control and experiment groups to diverge without a miscompilation being present.
\fuzzer{} statically rejects generations that combine multiple side-effecting sub-expressions within a single expression.

\paragraph{Environment-dependent instructions}

EVM instructions such as \texttt{gas}, \texttt{pc}, and \texttt{codesize} return values that legitimately differ between the control and experiment groups: optimized code consumes less gas, has different program counters, and produces smaller byte code.
Comparing the results of such instructions would flag divergences that are not miscompilations.
\fuzzer{} therefore excludes the results of such environment-dependent instructions from the compared runtime indicators.

\subsection{Design Trade-offs}

\paragraph{No ground truth}

Differential testing does not require a specification of correct compilation, but it can only detect divergence, not absolute correctness.
A miscompilation that manifests identically in both the control and experiment groups---for instance, a bug in the shared front end---goes unnoticed.
We accept this blind spot in exchange for a fully automated test oracle.

\paragraph{No guarantee of termination}

\fuzzer{} does not statically guarantee that generated programs terminate; for example, generated loops may fail to make progress.
We rely on the EVM's gas metering to bound execution: every execution is supplied a fixed gas budget, and an execution that exhausts its budget is terminated by the EVM.
Since gas exhaustion manifests identically in both groups, non-terminating programs do not produce false alarms; they merely test the compiler less effectively.

\paragraph{Target miscompilation bugs}

\fuzzer{} is designed to find miscompilation bugs rather than compiler crashes.
Design decisions such as generating well-typed programs reduce the diversity of inputs presented to the compiler front end, which is where crash-inducing inputs are typically rejected.
Nonetheless, internal compiler errors are detected as a by-product of testing and constitute a majority of the bugs found by \fuzzer{} (see Section~\ref{sec:quants}).

\section{Results and Discussion}
\subsection{Where Are The Bugs?}
\label{sec:rootcausedist}

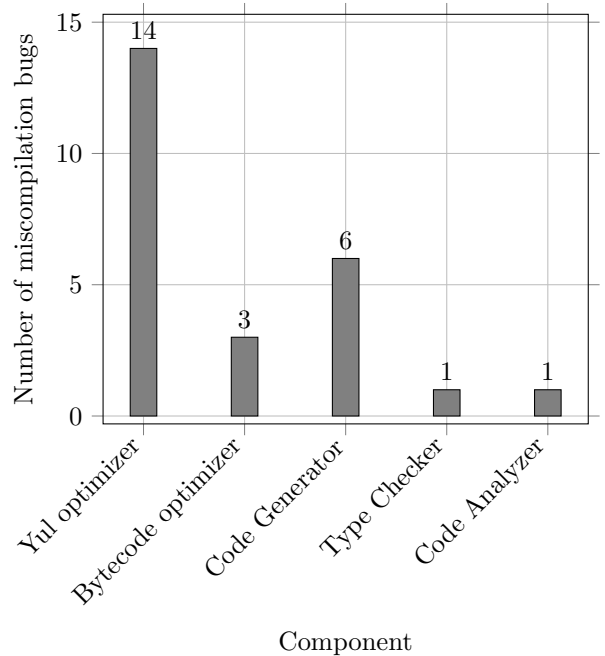
\begin{figure}
\begin{tikzpicture}
    \begin{axis}[
        width=8cm, height=7cm,     
        grid = major,
        axis background/.style = {fill=white},
        ylabel = {Number of miscompilation bugs},
        xlabel = {Component},
        tick align = outside,
        x tick label style={rotate=45, anchor=east, align=left},
	xticklabels = {Yul optimizer, Bytecode optimizer, Code Generator, Type Checker, Code Analyzer},
        xtick = {0,...,4},
	ybar,
	nodes near coords,
        ]
        \addplot[black, fill=gray] table[x expr=\coordindex, y=count] {
		x count
		YulOptimizer 14
		BytecodeOptimizer 3
		CodeGenerator 6
		TypeChecker 1
		CodeAnalyzer 1
	};
    \end{axis}
\end{tikzpicture}
\caption{Distribution of miscompilation bugs found by \fuzzer{} across compiler components.}
\label{fig:componentdist}
\end{figure}

Table~\ref{tbl:buglist} characterizes the miscompilation bugs found by \fuzzer{}.
Figure~\ref{fig:componentdist} shows their distribution across compiler components: over two thirds (17 out of 25) are optimizer bugs, with the Yul optimizer alone accounting for 14.
Table~\ref{tbl:bugdist} breaks the bugs down by root cause.
The most common root cause is an insufficient safety check, i.e., an optimization that is correct on most inputs but is applied to inputs on which it is not.
Table~\ref{tbl:filedist} shows where the bugs are located in the compiler code base: the single buggiest file is the list of algebraic optimization rules (\texttt{RuleList.h}) shared by the Yul and byte code optimizers, followed by the store eliminator optimization steps.

\begin{table*}
\small
\begin{tabular}{l | p{3cm} | p{6cm} | p{3cm}}
  Component & Bug location & Bug description & Impact \\
  \hline
  Yul optimizer & Optimizer rule & Side-effects due to expression re-ordering not accounted for (\href{https://github.com/ethereum/solidity/issues/7411}{\#7411}) & Incorrect computation \\
  Yul optimizer & Expression simplifier & Variable reassignments not accounted for (\href{https://github.com/ethereum/solidity/issues/6127}{\#6127})& Incorrect computation \\
  Yul optimizer & Redundant assignment eliminator & Side-effects of assignments that access memory not accounted for (\href{https://github.com/ethereum/solidity/issues/6827}{\#6827}) & Incorrect computation \\
  Yul optimizer & Dead-code eliminator & Dead code not properly removed (\href{https://github.com/ethereum/solidity/issues/6492}{\#6492}) & Syntactically incorrect transformation \\
  Legacy optimizer & Optimizer rule & Parameter ordering in rule swapped (\href{https://github.com/ethereum/solidity/issues/6316}{\#6316}) & Incorrect computation \\
  Legacy optimizer & Optimizer rule & Overflow check missing (\href{https://github.com/ethereum/solidity/issues/6246}{\#6246}) & Incorrect computation \\
  Legacy optimizer & Optimizer rule & Side-effects of expression unaccounted for (\href{https://github.com/ethereum/solidity/issues/7098}{\#7098}) & Incorrect computation \\
  Yul optimizer & Structural simplifier & Undefined behavior because of duplicate switch case expressions (\href{https://github.com/ethereum/solidity/issues/6359}{\#6359}) & Undefined behavior \\
  Legacy optimizer & Optimizer rule & Side-effect of expression unaccounted for (\href{https://github.com/ethereum/solidity/issues/9558}{\#9558}) & Incorrect computation \\
  Yul optimizer & Redundant store eliminator & EVM specification of opcode not respected (\href{https://github.com/ethereum/solidity/issues/13039}{\#13039}) & Incorrect control-flow \\
  Code generator & Yul IR generator & Missing clean-up post casting down to smaller fixed bytes type (\href{https://github.com/ethereum/solidity/issues/12535}{\#12535}) & Incorrect data \\
  Code generator & Yul IR generator & Incorrect forwarding of modifier input parameters (\href{https://github.com/ethereum/solidity/issues/12061}{\#12061}) & Incorrect code \\
  Code generator & Yul IR generator & Missing truncation before shift operation (\href{https://github.com/ethereum/solidity/issues/11736}{\#11736}) & Incorrect data \\
  Code generator & Yul IR generator & Incorrect function forwarding (\href{https://github.com/ethereum/solidity/issues/11631}{\#11631}) & Incorrect control-flow \\
  Code generator & Legacy optimizer & Incorrect re-use of cached keccak256 hash value (\href{https://github.com/ethereum/solidity/issues/11131}{\#11131}) & Incorrect computation \\
  Code generator & Legacy back-end & Missing clean-up of typed data (\href{https://github.com/ethereum/solidity/issues/11602}{\#11602}) & Incorrect data \\
  Yul optimizer & Optimizer rule & Incorrect transformation function (\href{https://github.com/ethereum/solidity/pull/9546}{\#9546}) & Incorrect computation \\
  Yul optimizer & Redundant store eliminator & Incorrect control-flow analysis (\href{https://github.com/ethereum/solidity/pull/11352}{\#11352}) & Incorrect computation \\
  Yul optimizer & Redundant store eliminator & Incorrect data-flow analysis (\href{https://github.com/ethereum/solidity/pull/12672}{\#12672}) & Incorrect computation \\
  Yul optimizer & Loop-invariant mover & Side-effect of expression not accounted for (\href{https://github.com/ethereum/solidity/issues/7847}{\#7847}) & Infinite loop \\
  Yul optimizer & Redundant assignment eliminator & Incorrect data-flow analysis (\href{https://github.com/ethereum/solidity/issues/8072}{\#8072}) & Incorrect computation \\
  Yul optimizer & Load Resolver & Incorrect control-flow analysis (\href{https://github.com/ethereum/solidity/issues/8032}{\#8032}) & Incorrect computation \\
  Front end & Type Checker & Permissive free function definition (\href{https://github.com/ethereum/solidity/issues/9851}{\#9851}) & Undefined behavior \\
  Yul optimizer & Common subexpression eliminator & Incorrect code transformation (\href{https://github.com/ethereum/solidity/issues/9308}{\#9308}) & Compiler error \\
  Yul optimizer & Redundant store eliminator & Incorrect removal of store (\href{https://github.com/ethereum/solidity/issues/13478}{\#13478}) & Incorrect state \\
\end{tabular}
\caption{Miscompilation bugs found by \fuzzer{}.}
\label{tbl:buglist}
\end{table*}

\begin{table}
\begin{tabular}{l | r}
Bug root cause & Number \\
\hline
Insufficient safety check & 11 \\
Incorrect analysis & 4 \\
Lazy clean-up & 3 \\
Implementation flaw & 3 \\
Overly permissive parsing & 3 \\
Incorrect optimization & 1 \\
\hline
Total & 25 \\
\end{tabular}
\caption{Distribution of the root cause of bugs found by \fuzzer{}.}
\label{tbl:bugdist}
\end{table}

\begin{table}
\begin{tabular}{p{2.5cm} | p{2cm} | r}
File Name & Purpose & Num. Bugs \\
\hline
RuleList.h & List of optimization rules & 6 \\
\seqsplit{UnusedStoreEliminator.cpp} & Optimization pass & 4 \\
\seqsplit{YulUtilFunctions.cpp} & Solidity to Yul utilities & 2 \\
Misc. Yul optimization modules & Yul optimization passes & 6 \\
Misc. legacy optimization modules & Byte code optimization passes & 3 \\
Misc. Solidity to Yul modules & Code transformation & 2 \\
Misc. analysis modules & Semantic code analysis & 2 \\
\hline
\end{tabular}
\caption{Distribution of miscompilation bugs found by \fuzzer{} across compiler source files.}
\label{tbl:filedist}
\end{table}

\subsection{Examples of Miscompilation Bugs}
\label{para:quals}

\paragraph{Bug 1: incorrect Keccak256 caching} (\href{https://github.com/ethereum/solidity/issues/11131}{\#11131})
The \texttt{keccak256(p, s)} hash function is a first-class opcode that computes the hash value of EVM memory region in the byte range \texttt{[p, p + s]}, where \texttt{p, s} (start pointer, length) are non-negative integers.
We found that the Solidity compiler since its very first release until version \texttt{0.8.2} contained a bug that would result in the hash values of overlapping memory regions in successive calls to the keccak function being incorrectly computed.
The problem occurred when a call to \texttt{keccak256(p, s)} was followed by another call \texttt{keccak256(p, s')} such that \texttt{mod(s, 32) == 0}, \texttt{mod(s', 32) != 0} (the first but not the second length parameter is a multiple of 32) and \texttt{s' < s} (the memory region for the second keccak call was contained within the first memory region).
The root-cause of the problem was that the Solidity legacy optimizer wrongly rounded up the \texttt{length} argument of the keccak function to the nearest multiple of 32 and in doing so, considered \texttt{keccak256(p, s) == keccak256(p, s')} although \texttt{s != s'}.
This happened because of two reasons: (1) the mistaken assumption that keccak hashes are always computed over memory regions that span a multiple of 32 bytes; (2) keccak256 hash computation of memory regions that may deduced to be the same (i.e., identical start pointer and length rounded upto 32) may be optimized by performing the computation for the first call, caching the computed value, and re-using the cached value for the second call instead of computing it again.
The wrong deduction caused the keccak256 hash value computed by line 5 of Listing~\ref{lst:keccakbug} and subsequently cached to be re-used on line 6, eventually returning \texttt{true} instead of the correct return value \texttt{false}.

\begin{lstlisting}[caption=Multiple versions of the Solidity compiler miscompiled this function wrongly returning true. The correct return value is false., label=lst:keccakbug]
contract C {
  function f() public returns (bool ret) {
    assembly {
      mstore(0, 0)
      let a := keccak256(0, 32)
      let b := keccak256(0, 23)
      ret := eq(a, b)
    }
  }
}
\end{lstlisting}

\paragraph{Bug 2: unaccounted side-effects} (\href{https://github.com/ethereum/solidity/issues/7411}{\#7411})
The Yul optimizer encodes an optimization rule that simplifies \texttt{MUL(X, SHL(Y, 1))} to \texttt{SHL(Y, X)}: an expression \texttt{X} multiplied by one left-shifted expression \texttt{Y} times is equal to \texttt{X} left-shifted \texttt{Y} times, eliminating the multiplication.
We found a bug in the Yul optimizer shipped with Solidity compiler versions prior to \texttt{0.5.12} that would result in functional expressions with side-effects (functions that return a single output and modify shared state) being incorrectly computed.
The root-cause of the problem lay in the swapped evaluation order of parameters due to the optimization rule.
This happened because in the original code \texttt{MUL(X, SHL(Y, 1))}, the evaluation order in Yul being left-to-right, the expression \texttt{X} was computed before \texttt{Y}, but vice versa in the optimized code \texttt{SHL(Y, X)}.
Therefore, the result on line 14 of Listing~\ref{lst:mulshlbug} is incorrectly computed as two because the function \texttt{readValue()} is evaluated before the function \texttt{writeValue()}, instead of the other way round as intended in the unoptimized code that would result in the value eight.
Although the optimization rule is computationally correct, it does not account for side-effects of expressions that depend on the evaluation order.
Interestingly, this optimization rule was proven to be correct using Z3 at the time the bug was found.
This is not surprising since the proof does not account for side-effects.

\begin{lstlisting}[caption=The Yul optimizer incorrectly optimizes the function bug() to return two. The correct return value is eight., label=lst:mulshlbug]
  {
    function readValue() -> x
    {
      x := sload(0)
    }
    function writeValue() -> y
    {
      sstore(0, 2)
      y := sload(0)
    }
    function bug() -> z
    {
      // Post optimization: z := shl(readValue(), writeValue())
      z := mul(writeValue(), shl(readValue(), 1))
    }
  }
\end{lstlisting}

\paragraph{Bug 3: Incorrect removal of storage writes} (\href{https://github.com/ethereum/solidity/issues/13478}{\#13478})
The Yul optimizer contains an optimization pass called unused store eliminator that removes provably redundant writes to storage and memory.
We found a bug in the optimization pass shipped with Solidity versions \texttt{0.8.13}--\texttt{0.8.16} that would lead to an incorrect state of persistent program storage in programs containing function calls that would write to storage and conditionally terminate.
The root-cause of the bug lay in not summarizing the effect of a storage write in conditionally terminating functions.

The Yul program in Listing~\ref{lst:storeremovebug} demonstrates the incorrect storage removal bug.
The function~\texttt{f()} writes a one to storage slot zero and then calls the function~\texttt{g()}.
The latter in turn gracefully returns the function via the~\texttt{leave} statement in case the 32-byte value at memory location zero is two.
Otherwise, program execution is terminated (i.e., the top-level transaction is terminated) via the~\texttt{return} statement.
The top-level program makes two calls to the function~\texttt{f()}.
The compiler generates code that incorrectly removes the first write to storage (via the first call to~\texttt{f()}).
Due to this removal, program persistent storage is empty should the call to~\texttt{g()} via the first top-level call to~\texttt{f()} terminate.
The correct state of storage is a one written to slot zero and the compiler must therefore retain the first storage write.
The fix for this bug is to add the additional safety check while annotating store statements: if a function may terminate, storage must be retained.

\begin{lstlisting}[caption=The Yul optimizer incorrectly removes the first storage write to slot zero resulting in an incorrect blockchain state after conditional program termination., label=lst:storeremovebug]
{
    function f() {
        sstore(0, 1)
        g()
    }
    function g() {
        switch mload(0)
        case 2 { leave }
	// terminate execution
        return(0, 0)
    }
    f()
    f()
}
\end{lstlisting}

\subsection{Longitudinal Analysis}
\label{sec:quants}

We performed a longitudinal analysis of the Solidity compiler using tests generated by \fuzzer{}.
We tested Solidity compiler releases that were made in the past three years and recorded the bugs that were discovered in each of them.
We use crash count---number of test cases that lead to an internal compiler error---to gauge the robustness of a release.
Since the discovered bugs have now been fixed, we analyze the nature of discovered bugs over the course of several years.

\paragraph{Crash count}

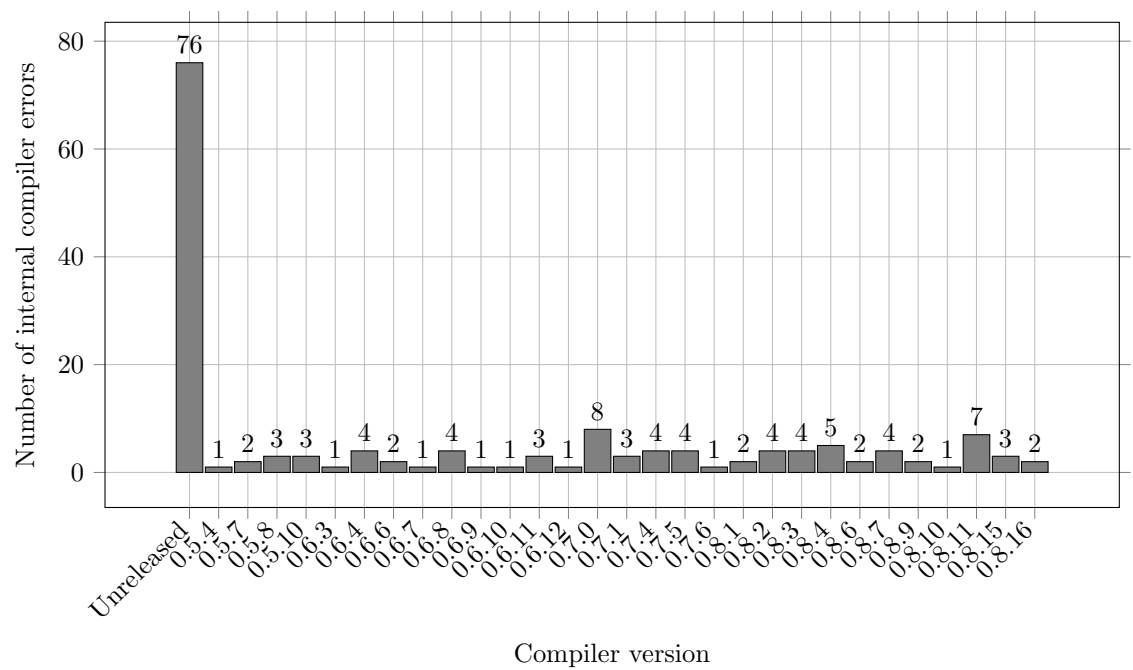
\begin{figure*}
\begin{tikzpicture}
    \begin{axis}[
        width=15cm, height=8cm,     
        grid = major,
        axis background/.style = {fill=white},
        ylabel = {Number of internal compiler errors},
        xlabel = {Compiler version},
        tick align = outside,
        x tick label style={rotate=45, anchor=east, align=left},
	xticklabels = {Unreleased, 0.5.4, 0.5.7, 0.5.8, 0.5.10, 0.6.3, 0.6.4, 0.6.6, 0.6.7, 0.6.8, 0.6.9, 0.6.10, 0.6.11, 0.6.12, 0.7.0, 0.7.1, 0.7.4, 0.7.5, 0.7.6, 0.8.1, 0.8.2, 0.8.3, 0.8.4, 0.8.6, 0.8.7, 0.8.9, 0.8.10, 0.8.11, 0.8.15, 0.8.16},
        xtick = {0,...,29},
	ybar,
	nodes near coords
        ]
        \addplot[black, fill=gray] table[x expr=\coordindex, y=count] {
		x count
		Unreleased 76
		0.5.4 1
		0.5.7 2
		0.5.8 3
		0.5.10 3 
		0.6.3 1 
		0.6.4 4 
		0.6.6 2 
		0.6.7 1 
		0.6.8 4 
		0.6.9 1 
		0.6.10 1 
		0.6.11 3 
		0.6.12 1 
		0.7.0 8 
		0.7.1 3 
		0.7.4 4 
		0.7.5 4 
		0.7.6 1 
		0.8.1 2 
		0.8.2 4 
		0.8.3 4 
		0.8.4 5 
		0.8.6 2 
		0.8.7 4 
		0.8.9 2 
		0.8.10 1 
		0.8.11 7 
		0.8.15 3 
		0.8.16 2
	};
    \end{axis}
\end{tikzpicture}
\caption{Number of internal compiler errors found by \fuzzer{} per Solidity compiler version. Errors labeled ``Unreleased'' were found and fixed before they shipped in a release.}
\label{fig:crashdist}
\end{figure*}

\begin{figure}
\begin{tikzpicture}
    \begin{axis}[
        width=8cm, height=7cm,     
        grid = major,
        axis background/.style = {fill=white},
        ylabel = {Number of internal compiler errors},
        xlabel = {Component},
        tick align = outside,
        x tick label style={rotate=45, anchor=east, align=left},
	xticklabels = {SMTChecker, TypeChecker, IRGenerator, BytecodeGenerator, SemanticAnalyzer,  Optimizer, Miscellaneous},
        xtick = {0,...,6},
	ybar,
	nodes near coords,
        ]
        \addplot[black, fill=gray] table[x expr=\coordindex, y=count] {
		x count
		SMTChecker 39
		TypeChecker 37
		IRGenerator 23
		BytecodeGenerator 22
		SemanticAnalyzer 21
		Optimizer 9
		Miscellaneous 13
	};
    \end{axis}
\end{tikzpicture}
\caption{Distribution of internal compiler errors found by \fuzzer{} across compiler components.}
\label{fig:icecomponents}
\end{figure}
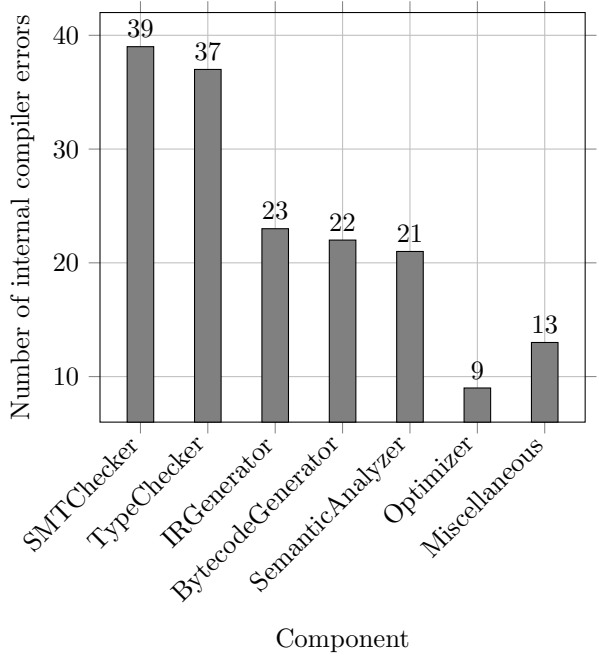

In addition to the 25 miscompilation bugs, \fuzzer{} has found 164 internal compiler errors: inputs on which the compiler terminates abnormally instead of producing output or a diagnostic.
Figure~\ref{fig:icecomponents} shows their distribution across compiler components.
The SMT-based model checker and the type checker together account for close to half of the internal compiler errors; both are front end components that every generated program exercises.
Figure~\ref{fig:crashdist} shows the compiler version in which each internal compiler error was first observed.
Notably, the largest single share of internal compiler errors (76) was found in unreleased code, i.e., the defect was caught by \fuzzer{} before it shipped in a compiler release.

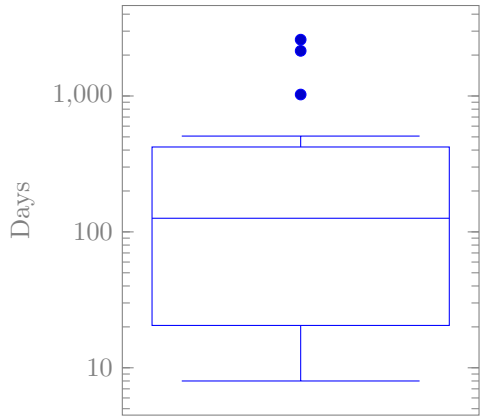
\begin{figure}
\begin{tikzpicture}
	\begin{axis}[
		width=6.3cm, height=7cm,
		boxplot/draw direction=y,
		color=gray,
		xmajorticks=false,
		ymode=log,
		log ticks with fixed point,
		ylabel={Days},
	]
		\addplot+[boxplot]
		table[row sep=\\,y index=0] {
			data\\
			126\\ 387\\ 501\\ 8\\ 15\\ 15\\ 69\\ 507\\ 112\\ 12\\ 14\\ 27\\ 74\\ 1025\\ 315\\ 456\\ 262\\ 2148\\ 2597\\ 26\\ 141\\ 176\\
			};
	\end{axis}
\end{tikzpicture}
\caption{Number of days a miscompilation bug found by \fuzzer{} persisted in the code base (log scale).}
\label{fig:bugdays}
\end{figure}

\begin{figure}
\begin{tikzpicture}
	\begin{axis}[
		width=7cm, height=7cm,
		boxplot/draw direction=y,
		xmajorticks=false,
		ymode=log,
		log ticks with fixed point,
		ylabel={Days},
	]
		\addplot+[boxplot]
		table[row sep=\\,y index=0] {
			data\\
			15\\ 15\\ 12\\ 14\\ 2148\\ 2597\\ 26\\ 176\\
			};
	\end{axis}
\end{tikzpicture}
\caption{Number of days a severity-rated miscompilation bug found by \fuzzer{} persisted in the code base (log scale).}
\label{fig:criticaldays}
\end{figure}
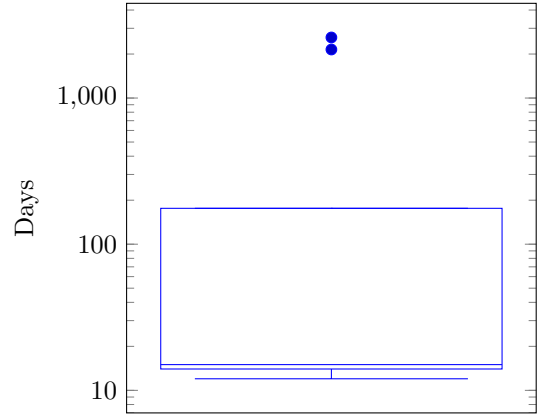

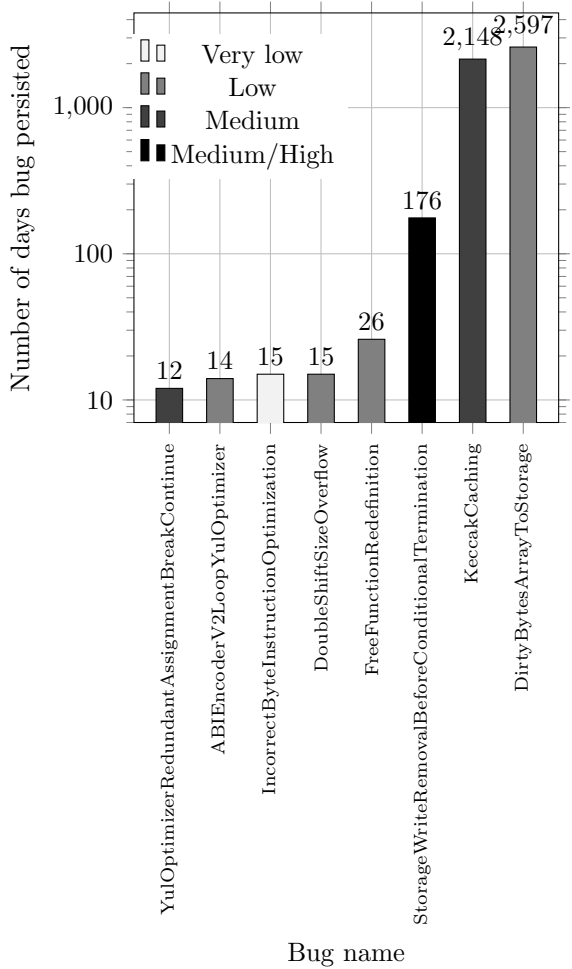
\begin{figure}
\begin{tikzpicture}
	\begin{axis}[
	        width=7.2cm, height=7cm,     
	        grid = major,
	        axis background/.style = {fill=white},
	        ylabel = {Number of days bug persisted},
	        xlabel = {Bug name},
		ymode = log,
		log ticks with fixed point,
	        tick align = outside,
	        x tick label style={rotate=90, anchor=east, align=left, font=\scriptsize},
		xticklabels = {YulOptimizerRedundantAssignmentBreakContinue, ABIEncoderV2LoopYulOptimizer, IncorrectByteInstructionOptimization, DoubleShiftSizeOverflow, FreeFunctionRedefinition, StorageWriteRemovalBeforeConditionalTermination, KeccakCaching, DirtyBytesArrayToStorage},
	        xtick = {0,...,7},
		ybar,
		nodes near coords,
		point meta=rawy,
	        every axis plot/.append style={
			ybar,
			bar shift=0pt,
			fill
		},
		legend style={draw=none,/tikz/every even column/.append style={column sep=0cm}},
		legend style={at={(0.25,0.95)},anchor=north},
	]
	\addplot[black, fill=black!5] coordinates{(2, 15)};
	\addplot[black, fill=black!50] coordinates{(1, 14) (3, 15) (4, 26) (7, 2597)};
	\addplot[black, fill=black!75] coordinates{(0, 12) (6, 2148)};
	\addplot[black, fill=black] coordinates{(5, 176)};
	\legend{Very low, Low, Medium, Medium/High}

	\end{axis}
\end{tikzpicture}
\caption{Number of days each severity-rated miscompilation bug found by \fuzzer{} persisted in the code base (log scale), shaded by assigned severity.}
\label{fig:severitydays}
\end{figure}

\paragraph{Bug persistence}

Figure~\ref{fig:bugdays} shows how long the miscompilation bugs found by \fuzzer{} persisted in the code base, measured as the number of days between the change that introduced a bug and its fix, for the 22 bugs whose introducing change we could identify.
The median miscompilation bug persisted for 134 days; roughly a third (7 out of 22) persisted for over a year, and the two longest-lived bugs persisted for roughly six (KeccakCaching, 2148 days) and seven (DirtyBytesArrayToStorage, 2597 days) years respectively.
This shows that routine testing not only misses miscompilation bugs at the time they are introduced but is unlikely to catch them later: bugs stay hidden until either a fuzzer or---worse---an affected user finds them.

\paragraph{Bug severity}

Of the 25 miscompilation bugs found by \fuzzer{}, 8 affected a released compiler version in a manner that warranted a severity rating and a public security alert; the remaining 17 were fixed before they shipped in a release.
Figure~\ref{fig:criticaldays} and Figure~\ref{fig:severitydays} show how long these severity-rated bugs persisted in the code base.
None of the bugs was rated high or critical severity, chiefly because the affected code patterns were judged unlikely to occur in typical production contracts.
The highest-rated bug (StorageWriteRemovalBeforeConditionalTermination, rated medium/high) silently removed a storage write, corrupting persistent contract state.
Notably, severity is uncorrelated with how long a bug persisted: the two longest-lived bugs were rated medium and low respectively.

\section{Conclusion}
We created~\fuzzer{} to find correctness bugs in the Solidity compiler, the most popular compiler for smart contracts deployed in the public Ethereum blockchain.
Using~\fuzzer{}, we found over two dozen miscompilation bugs in the Solidity compiler.
These bugs are serious because they cause the compiler to emit incorrect code in a manner that alters runtime semantics.
Although the bugs are serious,~\fuzzer{} helped soften their impact by finding them early, sometimes during code review.
Only about a third (8 out of 25) of miscompilation bugs found were in production code and assigned a severity rating.
Our work demonstrates that it is possible to find miscompilation bugs before they endanger end-users.

To find miscompilation bugs, the key problem we solved was to create semantically valid programs that are likely to stress test security critical components of the compiler, while avoiding programs that are likely to result in false alarms or typing errors.
The key insight of our study is that miscompilations occur due to various reasons such as not anticipating safety checks sufficiently, lazy cleanup of data, and permissive parsing.

The cost-benefit analysis of differential fuzz testing for finding miscompilation bugs is promising.
The rental costs of the AWS compute instance that was used to find miscompilation bugs over the course of more than three years amount to a little under~\textdollar{}15000, suggesting an average computational cost per bug of under~\textdollar{}1000.
Differential fuzz testing is not only economically viable but also promotes secure software development by making it less likely that miscompilation bugs slip to production.
Security alerts were avoided for two thirds of the miscompilation bugs (17 out of 25) found by \fuzzer{} because they were found before production use.

\paragraph{Software}~\fuzzer{} is open source and is developed as part of the Solidity compiler.
It is available at~\url{https://github.com/ethereum/solidity}.

\bibliography{citations}

\begin{thebibliography}{1}

\bibitem{godefroid}
P.~Godefroid, A.~Kiezun, and M.~Y. Levin, ``Grammar-based whitebox fuzzing,''
  {\em SIGPLAN Not.}, vol.~43, p.~206–215, jun 2008.

\bibitem{jsfunfuzz}
J.~Ruderman, ``Fuzzing for javascript correctness.''
  \url{https://www.squarefree.com/2007/08/02/fuzzing-for-correctness/}, 2007.

\bibitem{csmith}
X.~Yang, Y.~Chen, E.~Eide, and J.~Regehr, ``Finding and understanding bugs in c
  compilers,'' in {\em Proceedings of the 32nd ACM SIGPLAN Conference on
  Programming Language Design and Implementation}, PLDI '11, (New York, NY,
  USA), p.~283–294, Association for Computing Machinery, 2011.

\bibitem{emi}
V.~Le, M.~Afshari, and Z.~Su, ``Compiler validation via equivalence modulo
  inputs,'' in {\em Proceedings of the 35th ACM SIGPLAN Conference on
  Programming Language Design and Implementation}, PLDI '14, (New York, NY,
  USA), p.~216–226, Association for Computing Machinery, 2014.

\bibitem{soliditydocs}
S.~Team, ``Solidity documentation.''
  \url{https://docs.soliditylang.org/en/latest/}, 2022.

\end{thebibliography}
\bibliographystyle{ieeetr}

\end{document}